\documentclass[12pt]{article}
 
\usepackage[]{natbib}
\usepackage[]{aasms4}
\usepackage[]{graphicx}
 
\begin{document}

\title{Using Satellites to Probe Extrasolar Planet Formation}

\author{Paul Withers$^{a,*}$, J.~W. Barnes$^{b}$}

\affil{$^{a}$ Center for Space Physics, Boston University, 725 Commonwealth Avenue,
Boston, MA 02215, USA.}

\affil{$^{b}$ Department of Physics, University of Idaho, 875 Perimeter Drive,
Moscow, ID 83844-0903, USA.}

\affil{$^{*}$ Corresponding author email address: withers@bu.edu}





\noindent
{\bf Editorial Correspondence to:} \\
Paul Withers \\
Center for Space Physics, Boston University, \\
725 Commonwealth Avenue, Boston, MA 02215, USA. \\
Email: withers@bu.edu \\
Phone: (617) 353 1531 \\
Fax: (617) 353 6463

\pagebreak

\begin{abstract}
Planetary satellites are an integral part of the heirarchy of planetary systems. Here we make two predictions concerning their formation. First, primordial satellites, which have an array of distinguishing characteristics, form only around giant planets. If true, the size and duration of a planetary system's protostellar nebula, as well as the location of its snow line, can be constrained by knowing which of its planets possess primordial satellites and which do not. Second, all satellites around terrestrial planets form by impacts. If true, this greatly enhances the constraints that can be placed on the history of terrestrial planets by their satellites' compositions, sizes, and dynamics.
\end{abstract}


\pagebreak

The formation of stars from gas and dust is a fundamental astrophysical process, yet stars do not form alone. A large and rapidly growing number of stars are known to be orbited by debris disks or planets.
A comprehensive understanding of the life cycle of protostellar nebulae requires consideration of all of the types of condensed objects that form within them, not just stars. This includes planets, satellites, and objects analogous to asteroids or Kuiper belt objects. 
Many studies have investigated how the properties of planets and belts of smaller objects can be used to explore the formation of the planetary system, yet satellites have been somewhat neglected. 

Here we explore how the existence and nature of a planet's satellite system can be used to constrain how the planet formed.  We make two predictions concerning the formation of planetary satellites and investigate their consequences. 
First, we predict that primordial satellites, which only form in dense regions of a protostellar nebula, are found only around giant planets. 
The distribution of primordial satellites in a planetary system can be used to constrain the location of a protostellar nebula's snow line and thereby to empirically constrain whether close-in extrasolar planets formed \emph{in-situ} or migrated to their present locations. 
Second, we predict that all satellites around terrestrial planets formed by impacts. 
The composition of all terrestrial planets and their satellites reflects mixing across a broad region of the protostellar nebula.
Rapid advances in observational capabilities suggest that the first satellite of an extrasolar planet will soon be discovered 
\citep{1999A&AS..134..553S,2009MNRAS.400..398K}.
Our predictions outline the potential significance of such discoveries for the histories and current states of these satellites, planets, and planetary systems.

All planets in our solar system can be classified as either terrestrial planets (Mercury, Venus, Earth and Mars) or giant planets (Jupiter, Saturn, Uranus and Neptune). The key physical characteristic that defines the boundary between these two classes is composition, not size. The abundance of volatile species relative to refractory species is much greater for giant planets than terrestrial planets. This difference is attributed to differences in how the planets formed. The giant planets formed in $\sim$10 Myr from protoplanetary disks with high surface densities of volatile species, whereas the terrestrial planets formed in $\sim$100 Myr from the accretion of refractory planetesimals 
\citep{lissauer1993, boss2002}.

Although recent observations of extrasolar planetary systems have drastically reshaped ideas of planet formation, models still predict a bimodal distribution of volatile-rich planets analogous to giant planets and refractory-rich planets analogous to terrestrial planets 
\citep{reipurth2007}.
Since satellites are abundant in the solar system, we expect that satellites are likely to be present around many extrasolar planets (as long as they have not been lost due to tidal evolution; see \citet{2002ApJ...575.1087B}). Several theories of satellite formation have been proposed to explain the diverse satellites within the solar system, including in situ formation in a protoplanetary disk, gravitational capture, atmospheric capture, fission, and the impact between a planet and another body
\citep{stevenson1986}.

While the basic mechanisms of satellite formation are still actively debated by planetary scientists, certain trends
are apparent. In particular, large, prograde satellites orbiting in the 
equatorial plane of giant planets all seem to have formed \emph{in-situ} within the planet-forming
sub-nebula \citep[\emph{e.g.}][]{2006Natur.441..834C}.  Highly inclined (even retrograde) satellites around giant
planets formed via a capture mechanism \citep{2008AJ....136.1463V,2009AJ....137.5003N}. Satellites of
large solid planets are all consistent with an impact origin \citep{2001Natur.412..708C,2005Sci...307..546C}. 
Some asteroid satellites formed by fission \citep{Scheeres2007370}.  These mechanisms are candidates for how
extrasolar satellites form.

Prediction 1: Satellites that formed in the same place and at the same time as their parent planet exist only around planets analogous to giant planets. 

This prediction results from the understanding that the formation of such primordial satellites requires locally high surface densities of gas, ice or dust close to the growing planet. They only form beyond the nebula's snow line.
These conditions occur in the protoplanetary disks within which giant planets, but not terrestrial planets, form 
\citep{stevenson1986, boss2002}.  Thus if primordial satellites were found around a close-in planet with an
orbital semimajor axis inside its star's snow line, the existence of the satellite would imply that the planet
formed further out and migrated inward rather than having formed in place.

The distinguishing characteristics of such satellites are: low eccentricity, prograde orbits near the planet's equatorial plane and well within the planet's Hill sphere; same age of formation as the parent planet; and elemental and isotopic compositions that, although potentially modified by accretion and subsequent processes, are related to the environmental conditions at the location and time of the formation of the parent planet.

Extrasolar planets with satellites that possess these characteristics are predicted to be volatile-rich. That provides a constraint on relationships between the planet's mass, density, size, temperature and spectrum. They are also predicted to have formed rapidly within a protoplanetary disk. Since their elemental and isotopic composition reflects conditions where they formed, trends in these compositions can be used to determine spatial variations in the protostellar nebula and to constrain planetary migration post-formation. Mapping the distribution of such planets within a planetary system places constraints on the size and duration of the protostellar nebula from which the star, planets and satellites formed.

If this prediction withstands scrutiny, then consideration of a stronger version may be warranted. Specifically, that all planets that formed within protoplanetary disks originally possessed primordial satellites. This would imply that if such a planet no longer possesses primordial satellites, like Neptune, then they must have been removed somehow. The removal process is likely to have been a major event that affected the state and subsequent evolution of the planet, such as a strong gravitational interaction with another planet or a series of impacts.

If this prediction is disproven, then the presence of a protoplanetary disk during planet formation is not a discriminant between volatile-rich giant planets and refractory-rich terrestrial planets, which has substantial implications for planet formation.

Prediction 2: All satellites above a threshold mass around planets analogous to terrestrial planets formed by the accretion of ejecta from an impact. 

This prediction results from the elimination of other possible formation mechanisms. Prediction 1 implies that planets analogous
to terrestrial planets do not possess primordial satellites. Capture of a body by a planet via any mechanism requires energy loss 
\citep{burns1992}, 
which is impractical for bodies exceeding some threshold mass 
\citep{goldreich2002} 
that we do not quantify in this work. It is difficult for a large solid body to accumulate enough angular momentum to form satellites by fission without also accumulating so much energy that it is catastrophically disrupted 
\citep{peale1986, stern1997, Scheeres2007370}.
In addition, a rapidly rotating body is likely to lose angular momentum by shedding loose material before approaching the threshold for fission 
\citep{richardson2006}.
Only mechanisms based upon satellite formation from impact ejecta remain viable. 

The distinguishing characteristics of satellites formed from the accretion of impact ejecta are 
\citep{hartmann1986, stern1997, canup2000}:
low abundance of siderophile constituents relative to the parent planet; low relative abundance
of volatiles; large satellite to planet mass ratio; large ratio of the orbital angular momentum of the satellite to the total angular momentum of the system; and isotopic composition that, although potentially modified by subsequent processes, suggests a common origin for
the material in the planet and satellite.

Extrasolar planets with satellites that possess these characteristics are predicted to be refractory-rich. This provides a constraint on relationships between the planet's mass, density, size, temperature and spectrum. They are also predicted to have formed slowly from the accretion of refractory planetesimals. Due to mixing of planetesimals, the elemental and isotopic composition of such planets reflects conditions across a broad region of the protostellar nebula. Mapping the distribution of such planets within a planetary system places constraints on the size and duration of the protostellar nebula from which the star, planets and satellites formed.

If this prediction withstands scrutiny, then consideration of a stronger version may be warranted. Specifically, that all such satellites around a planet formed from a single impact. This would imply that any large impact which produces a new satellite disrupts any pre-existing satellites. The composition and dynamics of all of the planet's satellites would be linked by their common origin in a single impact event.

If this prediction is disproven, then either at least one seemingly impractical mechanism for satellite formation must occur more easily than is currently thought or an unsuspected mechanism for satellite formation exists. Either case has substantial implications for satellite formation.

In order to better constrain the formation of extrasolar planets, we have made two predictions concerning the formation of planetary satellites. Both are consistent with current knowledge of the solar system. The first prediction has broad consequences for how the current state of a planetary system can be used to constrain temporal and spatial variations of conditions in the protostellar nebula within which the star, planets, satellites and other components of the system formed. The second prediction highlights the importance of stochastic impacts as a process that affects not only the geophysical and geochemical states of the objects involved, but also the hierarchical structure of planetary systems.

A meaningful planetary classification scheme should be based on currently observable characteristics, yet also be
related to planetary formation and history. Planets in the mass range $1 \mathrm{M_\oplus} \leq M \leq 10 \mathrm{M_\oplus}$ could either be ``mini-Neptunes", like GJ1214b \citep{2009Natur.462..891C} or ``super-Earths" like CoRoT-7b \citep{2009A&A...506..287L}.  Future discoveries of planets in this mass range are likely to blur the currently clear boundary between terrestrial planets and giant planets. Consideration of a planet's retinue of satellites, particularly primordial satellites, may help fix the boundaries of planetary classification schemes.

\indent{\bf{Acknowledgments\\}}
JWB is supported by a grant from the NASA Exobiology program.

\newpage


\bibliographystyle{icarus}




\end{document}